\documentclass[11pt]{article}
\pdfoutput=1
\usepackage{amsmath}
\usepackage{amssymb}
\usepackage{graphicx,bbm,mathrsfs}
\usepackage{nicefrac}
\usepackage{slashed}
\usepackage{bbm}
\usepackage{geometry}
\usepackage{jheppub}

\usepackage{dcolumn}   
\usepackage{bm}        
\usepackage{graphicx,mathrsfs}
\usepackage{nicefrac}
\usepackage{multirow}
\usepackage{color}
\usepackage{mathtools}
\usepackage{makecell}

\newcommand{\be}{\begin{equation}}
\newcommand{\ee}{\end{equation}}
\newcommand{\bea}{\begin{eqnarray}}
\newcommand{\eea}{\end{eqnarray}}

\newcommand{\tr}{\operatorname{tr}}

\usepackage{wasysym}
\usepackage{hhline,colortbl}

\usepackage{titlesec}

\titleformat*{\section}{\Large\bfseries}
\titleformat*{\subsection}{\large\bfseries}
\titleformat*{\subsubsection}{\large\bfseries}
\titleformat*{\paragraph}{\large\bfseries}
\titleformat*{\subparagraph}{\large\bfseries}

\makeatletter
\newcommand*{\prodsym}{%
  \DOTSB
  \mathop{
    \mathchoice
      {\rlap{\kern.3em\rotatebox[origin=c]{-90}{}}{\prod}}
      {\vcenter{\rlap{\kern.2em\rotatebox[origin=c]{-90}{}}}{\prod}}
      {\sum}{\sum}
  }\slimits@
}
\makeatother

\renewcommand\vec{\bm}

\begin{document}

\vspace*{1.2cm}

\begin{center}

\thispagestyle{empty}
{\LARGE 

The Neutrino Casimir Force
 }\\[10mm]

\renewcommand{\thefootnote}{\fnsymbol{footnote}}

{\large  
Alexandria~Costantino$^{\,a}$ \footnote{acost007@ucr.edu },\, 
Sylvain~Fichet$^{\,b}$ \footnote{sfichet@caltech.edu }\,
}\\[10mm]

\end{center} 
\noindent
\quad\quad\quad\textit{$^a$ Department of Physics \& Astronomy, 
    	    University of  California, Riverside,}

\noindent \quad\quad\quad \textit{CA
92521} \\



\noindent
\quad\quad\quad \textit{$^b$ ICTP South American Institute for Fundamental Research  \& IFT-UNESP,}

\noindent \quad\quad\quad \textit{R. Dr. Bento Teobaldo Ferraz 271, S\~ao Paulo, Brazil
}

\addtocounter{footnote}{-2}

\vspace*{12mm}

\begin{center}
{  \bf  Abstract }
\end{center}

In the low energy effective theory of the weak interaction, a macroscopic force arises when pairs of neutrinos are exchanged. We calculate the neutrino Casimir force between plates, allowing for two different mass eigenstates within the loop. We also provide the general potential between point sources. We discuss the possibility of distinguishing whether neutrinos are Majorana or Dirac fermions using these quantum forces.

\newpage

\section{Introduction}
\label{se:intro}

There exists a great body of experimental evidence \cite{Fukuda:1998mi,Ahmad:2002jz,Araki:2004mb,Ahn:2006zza} to suggest that neutrinos undergo flavor oscillations and therefore have mass. 
A neutrino can be described as a 2-component fermion, and two distinct possibilities exist to generate its mass. One possibility is that neutrinos mass mix with an extra SM-singlet, in which case both can be described together in 4-component \textit{Dirac} fermions. Alternatively, a neutrino mass can arise from lepton number-violating mass insertions, in which case neutrinos can be described as self-conjugate 4-component \textit{Majorana} fermions.

\begin{figure}[t]
\centering
	\includegraphics[width=0.7\linewidth,trim={0cm 0cm 0cm 0cm},clip]{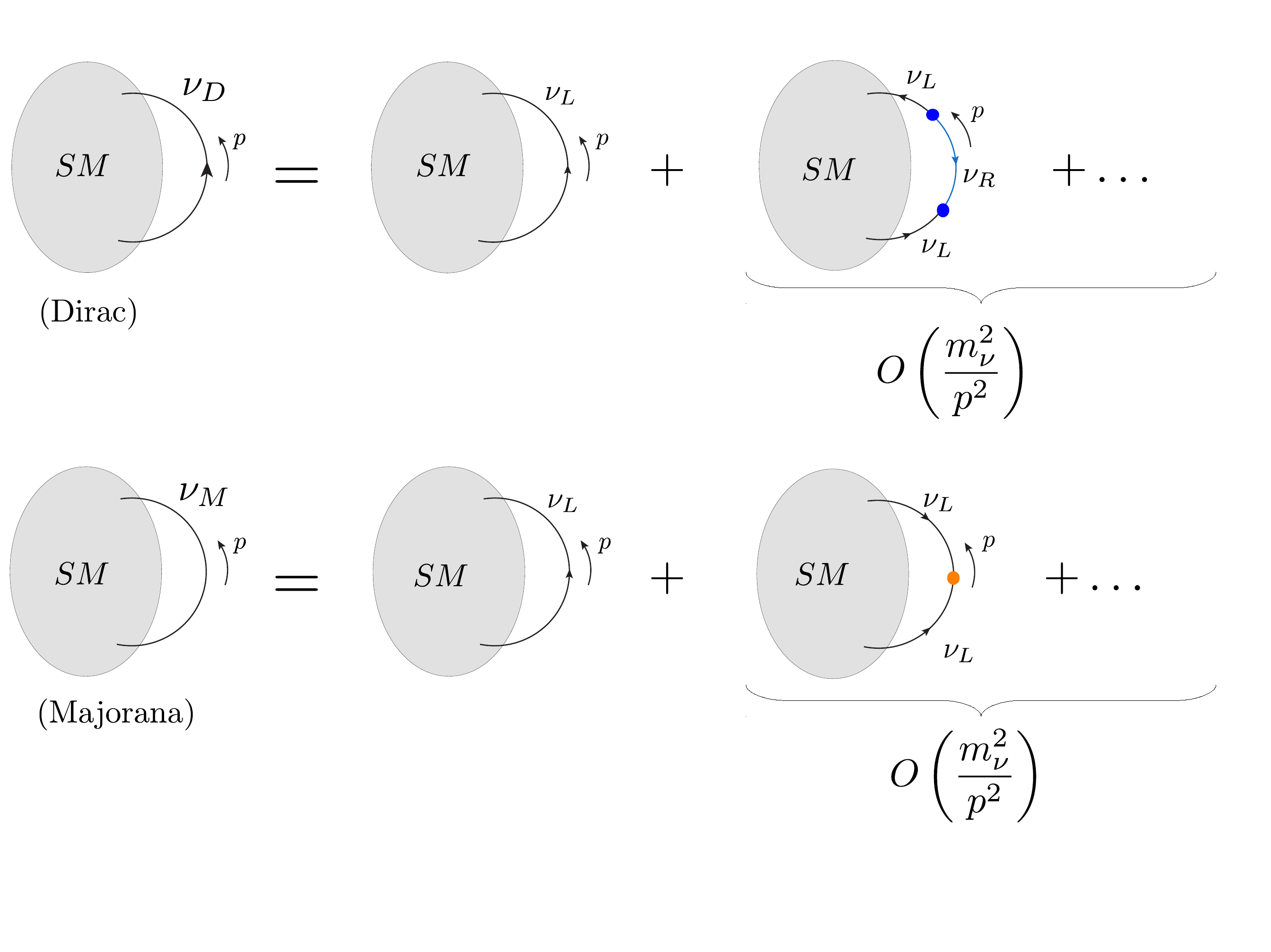}
	\caption{ The Majorana-Dirac confusion theorem. The blob represents an arbitrary SM amplitude from which we single out an internal neutrino propagator. Dirac mass insertions (top) and Majorana mass insertions (bottom) become negligible for $p\gg m_\nu$ such that amplitudes   become equivalent in this limit.
	}
	\label{fig:confusion}
\end{figure}

The difficulty in distinguishing these possibilities lies in the ``Majorana-Dirac confusion theorem''\cite{PhysRevD.26.1662,PhysRev.107.307}. 
In any amplitude, a neutrino propagator with 4-momentum  $p\gg m_\nu$  has mass insertions suppressed as $m^2_\nu/p^2$ and thus the mass generation mechanism cannot be observed. This is shown diagrammatically in Fig.\,\ref{fig:confusion}.
By unitary cuts the same property applies to external neutrino lines.\,\footnote{The confusion theorem is a property of the SM. In contrast, gravity  knows about all degrees of freedom and could identify whether an extra singlet neutrino exists, hence determining the nature of the neutrino mass. Existing approaches require one to consider the cosmological history of the Universe and depend on extra assumptions about physics beyond the SM  (see \textit{e.g.} \cite{Abazajian:2019oqj}). }

Massive neutrinos have a mass of order $0.1$\,eV\,(see \textit{e.g.} the upper bounds being placed by \cite{Aker:2019uuj}). This is much smaller than the energy scale of most typical scattering experiments, and therefore the confusion theorem makes the mass generation mechanism difficult to observe. 
In the laboratory, one approach has been to search for processes that are forbidden for Dirac neutrinos but are allowed for Majorana neutrinos. Such processes include the lepton number-violating neutrinoless double beta decay\cite{Agostini:2018tnm,Adams:2019jhp,KamLAND-Zen:2016pfg} and the neutrinoless double electron capture \cite{BERNABEU198315,PhysRevLett.106.052504,Bernabeu:2017ape}.

\begin{figure}
	\centering
	\includegraphics[width=0.3\linewidth]{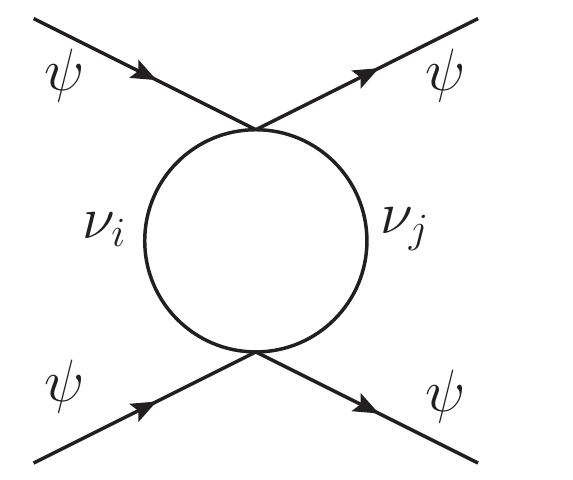}
	\caption{Quantum forces can be induced from the exchange of two neutrinos. The two virtual neutrinos can in principle have different masses $m_i\neq m_j$.}
	\label{fig:fd}
\end{figure}

Another approach is to study the macroscopic forces that arise from the exchange of virtual neutrinos \cite{Grifols:1996fk,Segarra:2020rah}. 
In the  low energy effective theory of the weak interaction, pairs of neutrinos can be exchanged, as shown in Fig.\,\ref{fig:fd}. This exchange results in long range \textit{quantum}\,\footnote{Quantum forces have their leading contribution at loop level. See \cite{Brax:2017xho,Fichet:2017bng,Costantino:2019ixl} for applications to dark sector searches.} forces between standard model fermions \cite{PhysRev.166.1638,FEINBERG198983,Hsu:1992tg,Thien:2019ayp,Stadnik:2017yge,Ghosh:2019dmi}. On distances the order of $1/m_\nu$ or larger the confusion theorem no longer presents any issue as the neutrino mass is not negligibly small at that length scale.  Hence by observing the potential, one could in principle distinguish the Dirac versus Majorana nature of neutrinos.

Recently there has been renewed interest in this approach, for instance in \cite{Segarra:2020rah}. These authors consider three generations of neutrinos (including mixing) in their calculation of the potential between point sources. They again find that a distinction between Majorana and Dirac neutrinos is possible when the separation of the point sources is on the order of $1/m_\nu$ or greater.

The sources considered in the previous works are pointlike. There also exists a force between extended macroscopic bodies: a neutrino Casimir force. 
In a realistic experimental setup aimed at establishing the nature of the neutrino masses, it is likely to be this Casimir force which is experimentally relevant, as the Compton wavelength of massive neutrinos is on the order of a micron, much larger than the atomic scale. 
In the case of planar geometry, the potential is dominated by long wavelength contributions and therefore it is not obvious how the confusion theorem applies. 
It is with these motivations that we study the neutrino Casimir force in the plate-plate and point-plate configurations. These can then serve as approximations of the force in more evolved geometries \cite{PFA}.

For completeness, we study the general point-point neutrino-induced quantum force in Sec. \ref{se:aa}. The evaluation uses the standard momentum-space formalism. We then introduce a mixed position-momentum space formalism and present the plate-plate and plate-point calculations 
in \ref{se:Casimir}. We discuss the results in Sec. \ref{se:con}.

\section{Potential Between Point Sources}\label{se:aa}


We use the 4-component fermion formalism in our loop calculations. At energies below the electroweak scale, the Lagrangian describing neutrino mass eigenstates interacting with SM fermions is given by
\be
\mathcal{L}_D= i\bar \nu^i_D \gamma^\mu \partial_\mu \nu^i_D - m_i \bar \nu^i_D\nu^i_D - \frac{G_F}{2\sqrt{2}}\left[\bar{\nu}_D^j\gamma^\mu\left(1-\gamma_5\right)\nu_D^i\right]\left[\bar{\psi}\gamma_\mu\left(g^V_{ij}-g^A_{ij}\gamma_5\right)\psi\right]\label{eq:lagrangian_D}
\ee
for Dirac neutrinos and 
\be
\mathcal{L}_M=\frac{i}{2}\bar \nu^i_M \gamma^\mu \partial_\mu \nu^i_M - \frac{m_i}{2} \bar \nu^i_M\nu^i_M + \frac{G_F}{2\sqrt{2}}\left[\bar{\nu}_M^j\gamma^\mu\gamma_5\nu_M^i\right]\left[\bar{\psi}\gamma_\mu\left(g^V_{ij}-g^A_{ij}\gamma_5\right)\psi\right]\label{eq:lagrangian_M}
\ee
for Majorana neutrinos. 
The link to the 2-component fermion notation is given in App.~\ref{app:fermions}. 
The $g^V_{ij}$ and $g^A_{ij}$ coupling matrices depend on the SM field and on the neutrino generation. They are given in App.~\ref{app:fermions} for completeness.

In this section, we present the force between two nonrelativistic fermions $\psi$ arising from the exchange of two neutrinos, $\nu_i$ and $\nu_j$. 
Similar results have already been presented in the literature, see for instance \cite{Grifols:1996fk,Segarra:2020rah}. Here we present the most complete result including the spin-dependent part of the potential. 

The calculation starts from the scattering amplitude in 4-momentum space. This formalism has been used in the literature (see \textit{e.g.} \cite{Ferrer:1998ue, Costantino:2019ixl} for details), so we only present results here. See App. \ref{se:aamath} for additional details.

We introduce a discrete variable to distinguish between the Majorana and Dirac cases:
\begin{align}
\eta &= 
\begin{cases}
0 & \text{if Majorana}\\
1 & \text{if Dirac}
\end{cases}
\ .
\end{align}
The full potential can be written using discontinuities (noted $\text{D}\left[f\right]$) across branch cuts of a basis of $f_{mn}$ functions
\begin{align}
f_{mn} &\equiv \int_{0}^{1}dx x^m(1-x)^n\ln \left(\frac{xm_j^2+(1-x)m_i^2 -x(1-x)\lambda^2}{\Lambda^2}\right),
\end{align}
which come from evaluating the loop integral. In this basis, the full potential is given by
\begin{align}
 & V(\bm r) =\sum_{ij}V_{ij}(\bm r) =  -iG_F^2\sum_{ij}\frac{\left(\delta_0^\mu\mathbbm{1}_1g^V_{ij}\delta_0^\nu\mathbbm{1}_2g^V_{ji}+\delta_c^\mu\sigma^c_1g^A_{ij}\delta_d^\nu\sigma^d_2g^A_{ji}\right) }{(4\pi)^2}\int_{m_i+m_j}^{\infty}\frac{\lambda d\lambda}{(2\pi)^2}\times \label{eq:aaintegral} 
\\ \nonumber
& \left(-g_{\mu\nu}\left(m_j^2\text{D}[f_{10}]+(1-\eta)m_im_j\text{D}[f_{00}]+m_i^2\text{D}[f_{01}]-2\lambda^2\text{D}[f_{11}]\right)+2\text{D}[f_{11}]\delta_\mu^a\delta_\nu^b \nabla_a\nabla_b \right)\frac{e^{-\lambda r}}{r}
\end{align}
where $\bm\sigma_{A}$ denotes the Pauli matrices acting on the spinors of source fermion $A$. Latin indices, such as $a,b,...$, are summed over $1,2,3$. We refer to the $V_{ij}$ as partial potentials. Summing over all combinations of neutrinos from the 3 generations yields the full quantum potential from neutrinos.

We can perform the integral exactly for the diagonal terms ($m_i=m_j\equiv m$). These partial potentials take on the form
\begin{align}
&V_{ii}(\bm r) = G_F^2\frac{\left(\delta_0^\mu\mathbbm{1}_1\left(g^V_{ii}\right)^2\delta_0^\nu\mathbbm{1}_2+\delta_c^\mu\sigma^c_1\left(g^A_{ii}\right)^2\delta_d^\nu\sigma^d_2\right)}{64\pi^4}\times\label{eq:diag}
\\& \nonumber \left(g_{\mu\nu}\left(\eta\frac{4m^3\pi K_3(2mr)}{r^2}+(1-\eta)\frac{8m^2\pi K_2(2mr)}{r^3}\right)+\delta_\mu^a\delta_\nu^b \nabla_a\nabla_b\frac{m\pi}{3r^2}\left(4K_1(2mr)+m\pi^2rG(m^2r^2)\right) \right)
\end{align}
where we have introduced the Meijer G-function
\begin{equation}
G(m^2r^2) \equiv G_{2,4}^{2,0}\left(m^2 r^2\bigg|
{\begin{array}{c}
	\frac{1}{2},\frac{3}{2} \\
	0,0,\frac{1}{2},\frac{1}{2} \\
	\end{array}}
\right) \,.
\end{equation}
The spin-independent piece of \eqref{eq:diag} is given by
\be
V_{ii}(r) = G_F^2\frac{\left(g^V_{ii}\right)^2}{16\pi^3}\left(\eta\frac{m^3 K_3(2mr)}{r^2}+(1-\eta)\frac{2m^2 K_2(2mr)}{r^3}\right),\label{eq:aapot}
\ee
which corresponds to a repulsive force and is consistent with the literature (\textit{e.g.} \cite{Grifols:1996fk, Segarra:2020rah}). At short distances $mr\ll1$, Dirac and Majorana predictions converge to 
\be
V_{ii}(r) = G_F^2\frac{\left(g^V_{ii}\right)^2}{16\pi^3r^5}\,,
\ee
as expected from the confusion theorem.

\section{The Neutrino Casimir Force }\label{se:Casimir}

Here we consider the quantum force  between extended sources. 
Focusing on nonrelativistic, unpolarized sources formed by the SM fermions, we have $\bar \psi \gamma_\mu \gamma_5 \psi\approx 0$, $\bar \psi \gamma_\mu \psi \approx \delta_{\mu 0} \psi^\dagger \psi= \delta_{\mu 0} \, n({\bm x})$ were $n({\bm x})$ is the number density operator. We denote by $J({\bm x})$ the  density expectation value  in the presence of matter, $J({\bm x})=\langle \Omega | n({\bm x}) | \Omega\rangle$. We can write  effective neutrino Lagrangians in the presence of such nonrelativistic static matter,
\be
\mathcal{L}_D=  i\bar \nu^i_D  \gamma^\mu \partial_\mu \nu^i_D - m_i \bar \nu^i_D\nu^i_D -  \frac{G_F}{2\sqrt{2}}\left[\bar{\nu}_D^j\gamma^0\left(1-\gamma_5\right)\nu_D^i\right] g^V_{ij}\,
J
\label{eq:lagrangian_D_ext}
\ee
\be
\mathcal{L}_M=\frac{i}{2}\bar \nu^i_M  \gamma^\mu \partial_\mu \nu^i_M - \frac{m_i}{2} \bar \nu^i_M\nu^i_M +  \frac{G_F}{2\sqrt{2}}\left[\bar{\nu}_M^j\gamma^0\gamma_5\nu_M^i\right] g^V_{ij}
\, J\label{eq:lagrangian_M_ext}\,.
\ee

We assume the matter density is compound of two pieces with density $J_1$, $J_2$, separated by a distance $L$. The full matter density is $J=J_1+J_2$. 
The potential between these two sources can be obtained by varying the quantum vacuum energy of the system with respect to $L$.

In case of strong coupling to sources, the neutrino would acquire an effective mass inside the sources, which tends to repel the propagators. This strong coupling regime reproduces precisely the familiar Casimir force with Dirichlet boundary conditions on the sources \cite{Brax:2018grq}. Calculations of forces in the strong coupling limit can be found  in \cite{milton2001casimir}. Instead, in case of weak coupling to sources, which is the one relevant here, the potential is given by the leading term of the one-loop functional determinant. 
 The force in this weakly coupled regime amounts to 
a ``Casimir-Polder'' force between extended objects. 
See App.~\ref{app:path} and Ref.\,\cite{Brax:2018grq} for details.
For simplicity, and because the  weak and strong regime of the force between extended objects have a unified description, we refer to the force in the weakly coupled regime as ``Casimir force''.

We find the  potential induced by the neutrinos between extended sources $J_1$ and $J_2$  to be \be
V(L) = i \frac{G_F^2}{2^\eta \, 4 } \int d^3 {\bm x} \int d^4 x'\, \sum_{ij} {\rm tr}\left[ J_1({\bm x}) \Delta_i(x,x') \,\Gamma\, g^V_{ij}
J_2({\bm x}') \Delta_j(x',x) \,\Gamma \, g^V_{ji} \right]+{\cal O} \left(G_F^3\right) \label{eq:Vgen}
\ee
where  $\Gamma=\gamma_0(\eta-\gamma_5)$ encodes the Lorentz structure of the neutrino vertex. 
Here $\Delta(x,x') $ is the Feynman propagator of 4-component fermions. The trace is in spinor space. 
Notice that one of the  integrals is in 3d space while the other is in spacetime. This reflects the fact that the quantum force is intrinsically relativistic. 

In the limit of pointlike sources 
\begin{align}
&J_1({\bm x})= \delta({\bm x}) &J_2({\bm x})=\delta({\bm x}-{\bm L}) ,
\end{align}
\eqref{eq:Vgen} reproduces \eqref{eq:potentialamplitude}
and thus 
the point-point potential obtained in Sec.~\ref{se:aa}.

\subsection{Potential Between Plates}\label{se:pp}

We consider the sources are infinite plates with separation $L$. The plates are taken to have number densities $n_1$ and $n_2$ and are orthogonal to the $z$ direction,
\begin{align}
&J_1({\bm x})=  n_1\Theta(z<0) &J_2({\bm x})=n_2\Theta(z>L)
\end{align}
The two transverse spatial coordinates are denoted by $x_\parallel$, hence $x_\mu=(t,{\bm x})=(t,x_\parallel,z)$. It is also useful to introduce the (2+1) Lorentz indexes $\alpha=0,1,2$ defining $x_\alpha=(t,x_\parallel)$. 

A naive method to obtain the plate-plate potential would be to directly integrate the general point-point result \eqref{eq:aaintegral}. This is however rather challenging in the case of different masses. We show here a simpler path to the general result.  

Since the sources are Lorentz-invariant along $x_\parallel$, we introduce  Fourier transforms along these coordinates and time. This introduces the 3-momentum conjugate of the $x_\alpha$ coordinates. 
In this mixed position-momentum space, the fermion propagators are found to be
\be
\Delta(k_\alpha,z-z')=\int\frac{dk_z}{2\pi}e^{ik_z(z-z')}\Delta(k_\mu)=\left(\slashed{k}+\omega_i\gamma^3\text{Sign}[z-z']+m\right)\frac{e^{i\omega_i\left|z-z'\right|}}{2\omega_i}
\ee
with 
\be
\slashed{k}=\gamma^\alpha k_\alpha\,, \quad \quad\omega_i\equiv\sqrt{k^2-m_i^2+i\varepsilon}\,,\quad \quad k^2=k_\alpha k^\alpha \,.\label{eq:prop_mixed}
\ee
Introducing the mixed space propagator in Eq.~\eqref{eq:Vgen} gives 
\begin{align}
V(L) = i \frac{G_F^2}{2^\eta \, 4 } & \int d^3 {\bm x}  \int d^4 x'
\int \frac{d^3k}{(2\pi)^3}
\int \frac{d^3k'}{(2\pi)^3}e^{i(k-k')_\alpha (x-x')^\alpha} \label{eq:Vgen1}
\, \\
&\sum_{ij} {\rm tr}\left[ J_1(z) \Delta_i(k_\alpha,z-z') \,\Gamma\, g^V_{ij}
J_2(z') \Delta_j(k'_\alpha,z'-z) \,\Gamma \, g^V_{ji} \right] \nonumber
\end{align}
A momentum redefinition makes appear the loop integral, the external momentum $q_\alpha$ and the overall Fourier transform in $q_\alpha$, 
\begin{align}
V(L) = i \frac{G_F^2}{2^\eta \, 4 } & \int d^3 {\bm x}  \int d^4 x'
\int \frac{d^3q}{(2\pi)^3}e^{iq_\alpha (x-x')^\alpha}
\int \frac{d^3k}{(2\pi)^3}
\label{eq:Vgen2}
\, \\
&\sum_{ij} {\rm tr}\left[ J_1(z) \Delta_i(k_\alpha,z-z') \,\Gamma\, g^V_{ij}
J_2(z') \Delta_j(k_\alpha+q_\alpha,z'-z) \,\Gamma \, g^V_{ji} \right] \nonumber
\end{align}

In the case of planar geometry considered here, it turns out that  the external 3-momentum is set to zero because of \be\int d^3 x \label{eq:delta} e^{iq_\alpha (x-x')^\alpha}=(2\pi)^3\delta^{(3)}\left(q_\alpha\right)\,.\ee
The fact that $q_0=0$ is a mere consequence of the nonrelativistic limit. The fact that $q_\parallel= 0$ is specific of the planar geometry and  indicates that the force is dominated  by fluctuations with infinite transverse wavelengths. 
The remaining transverse integral is factored as a surface $\int d^2 x_\parallel=S$, and the potential is given by 
\begin{align}
V(L) = i \frac{G_F^2}{2^\eta \, 4 }  S \int dz dz'
\int \frac{d^3k}{(2\pi)^3}
\sum_{ij} {\rm tr}\left[ J_1(z) \Delta_i(k_\alpha,z-z') \,\Gamma\, g^V_{ij}
J_2(z') \Delta_j(k_\alpha,z'-z) \,\Gamma \, g^V_{ji} \right] \label{eq:Vgen3}
\end{align}


Performing both remaining position integrals and evaluating the trace, we have
\be
V(L) = -i\frac{Sn_1n_2G_F^2}{4} \sum_{ij}g^V_{ij}g^V_{ji}\int \frac{d^3k}{(2\pi)^3} \left(\frac{2k_0^2 -k^2 -\omega_i \omega_j-(1-\eta)m_im_j}{\omega_i \omega_j(\omega_i+ \omega_j)^2 }\right)e^{i(\omega_i+ \omega_j)L}.
\ee
The only remaining integral is the loop integral. 
We Wick rotate the momentum integral from $2+1$ Lorentzian  to 3-dimensional Euclidian space,  
\be
V(L) = \frac{-Sn_1n_2G_F^2}{4} \sum_{ij}g^V_{ij}g^V_{ji} \int \frac{d^3k_E}{(2\pi)^3} \left(\frac{2k_{E0}^2-k_E^2-\omega_{Ei} \omega_{Ej}+(1-\eta)m_im_j}{\omega_{Ei} \omega_{Ej}(\omega_{Ei}+ \omega_{Ej})^2 }\right)e^{-(\omega_{Ei}+ \omega_{Ej})L}.
\ee
where we has defined $\omega_{Ei}=\sqrt{k^2_E+m^2_i}$.
We go to spherical coordinates and perform the angular integrals. For the remaining radial integral we introduce a dimensionless variable \be u=|k_E|L
\,,\quad \quad \quad \label{eq:defs}\rho_i=\sqrt{u^2+m_i^2 L^2}\,.
\ee
The potential between plates is found to be
 \be
V(L) = \frac{Sn_1n_2G_F^2}{8\pi^2 L} \sum_{ij}g^V_{ij}g^V_{ji} \int_0^\infty du u^2 \left(\frac{\frac{1}{3}u^2+\rho_i\rho_j -(1-\eta)m_im_jL^2}{\rho_{i} \rho_{j}(\rho_{i}+ \rho_{j})^2 }\right)e^{-(\rho_{i}+ \rho_{j})}.\label{eq:plplfinalpot}
\ee

The rest of the  integral  cannot be performed analytically in general. Notice the loop integral is finite   by construction because  the two sources have finite separation. In this calculation there is no need for any loop integral regularization, expressions  are finite at every step. 

The pressure between the plates is given by
\be
P(L)=\frac{-1}{S}\frac{\partial }{\partial L} V =
\frac{n_1n_2G_F^2}{8\pi^2 L} \sum_{ij}g^V_{ij}g^V_{ji} \int_0^\infty du u^2 \left(\frac{\frac{1}{3}u^2+\rho_i\rho_j -(1-\eta)m_im_jL^2}{\rho_{i} \rho_{j}(\rho_{i}+ \rho_{j}) }\right)e^{-(\rho_{i}+ \rho_{j})}.
\ee
The neutrino Casimir pressure is thus repulsive. 

Finally, at short distance \textit{i.e.} in the limit of $m_i,m_j\ll 1/L$, the integrals can be done exactly,
\begin{align}
V(L) = \frac{Sn_1n_2G_F^2}{48\pi^2L}\sum_{ij}g^V_{ij}g^V_{ji}& \quad \quad \quad P(L) = \frac{n_1n_2G_F^2}{48\pi^2L^2}\sum_{ij}g^V_{ij}g^V_{ji} .\label{eq:pppot2}
\end{align}
In this regime the Majorana and Dirac predictions have become equal, as expected from the confusion theorem.

\subsection{Potential Between a Plate and a Point Source}\label{se:ap}

To obtain the plate-point potential, we consider sources of the form

\begin{align}
&J_1({\bm x})=n_1\Theta(z>L)  &J_2({\bm x})=\delta({\bm x})
\end{align}
with \eqref{eq:Vgen2} and \eqref{eq:delta}. Performing the remaining position integrals and evaluating the trace, we have

\be
V(L) = \frac{-n_1G_F^2}{4} \sum_{ij} g^V_{ij}g^V_{ji}\int \frac{d^3k}{(2\pi)^3} \left(\frac{2k_0^2 -k^2-\omega_i \omega_j -(1-\eta)m_im_j}{\omega_i \omega_j(\omega_i+\omega_j) }\right) e^{i(\omega_i+\omega_j)L}.
\ee
We follow the steps of the plate-plate calculation---Wick rotating, performing the angular integral, and using the definitions \eqref{eq:defs}. We obtain 

\be
V(L) = \frac{n_1G_F^2}{8\pi^2L^2} \sum_{ij} g^V_{ij}g^V_{ji}\int_0^\infty du u^2 \left(\frac{\frac{1}{3}u^2+\rho_i \rho_j-(1-\eta)m_im_jL^2}{\rho_i \rho_j(\rho_i+ \rho_j) }\right)e^{-(\rho_i+ \rho_j)}.\label{eq:appot}
\ee
At short distances, $m_i,m_j\ll 1/L$ for all $(i,j)$, the integral can be done exactly, yielding
\be
V(L) = \frac{n_1G_F^2}{48\pi^2L^2} \sum_{ij} g^V_{ij}g^V_{ji}.\label{eq:appot2}
\ee
We again find that any trace of the mass generation mechanism has vanished from the short-distance result.

\section{Discussion}
\label{se:con}

The neutrino Casimir force has not previously been determined in the literature, to the best of our knowledge. In this section, we elucidate its properties.

The expressions for the neutrino Casimir force \eqref{eq:plplfinalpot}, \eqref{eq:appot} contain only one numerical integral, just as for the point-point result \eqref{eq:aaintegral}. This property generalizes to plates with an arbitrary number of layers, which is easily obtainable in our formalism. In our calculation, we take into account loops with two different mass eigenstates, that we denote below as $m_>=\text{max}\left(m_i,m_j\right), m_<=\text{min}\left(m_i,m_j\right)$.

\begin{figure}
	\centering
	\includegraphics[width=0.6\linewidth]{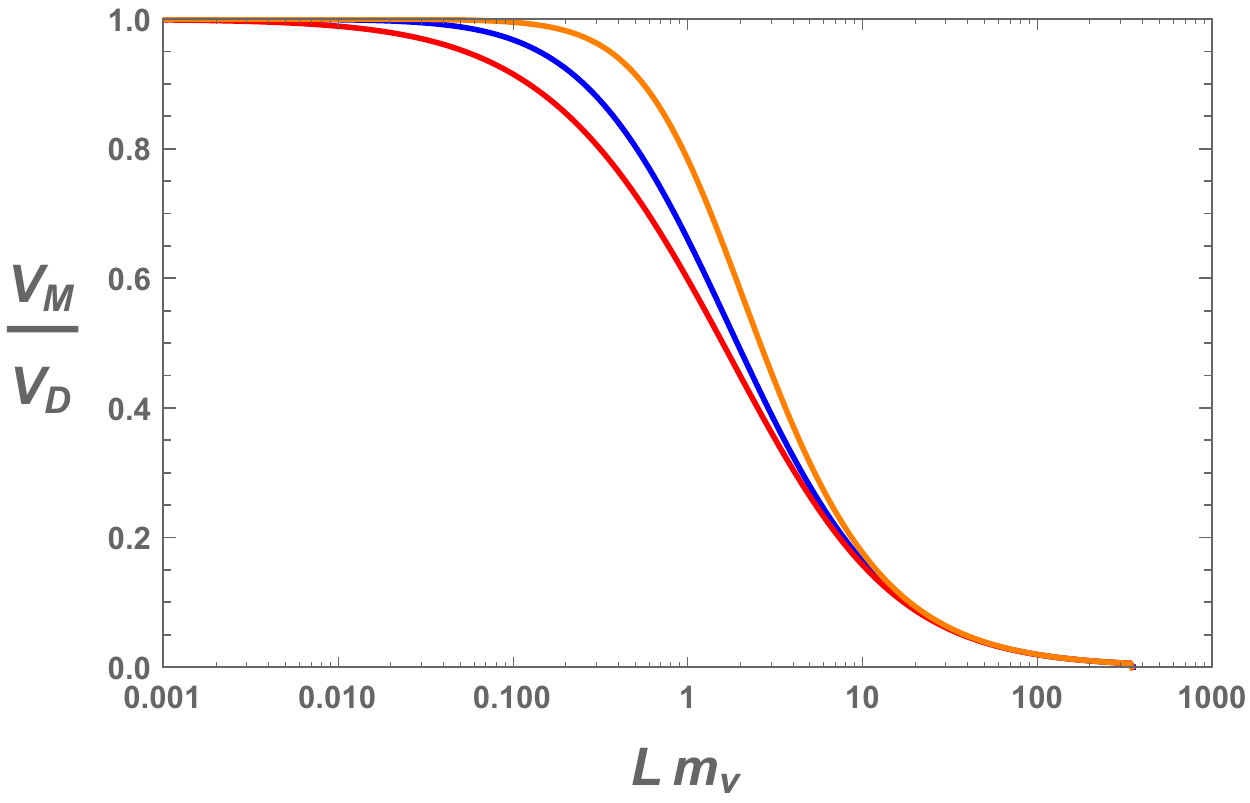}
	\caption{Ratios of the Dirac and Majorana potentials for point-point (orange), plate-point (blue), and plate-plate (red) configurations. Results are shown for equal masses ($m_i = m_j$). Majorana potentials are always weaker than Dirac potentials, consistent with prior results (\textit{e.g.} \cite{Grifols:1996fk}).}
	\label{fig:md-comparison}
\end{figure}

The Dirac and Majorana partial potentials $V_{ij}$ converge to each other in the limit of short distance, $L\ll 1/m_>$. The convergence holds for all configurations of sources considered and is shown for the case of equal masses ($m_i = m_j$) in Fig. \ref{fig:md-comparison}. This is the fingerprint of the confusion theorem---only the $\nu_L$ neutrino contributes to the pressure and thus any trace of the mass generation mechanism vanishes (see Fig.\,\ref{fig:confusion}). The plate-plate, plate-point, and point-point potentials scale as $1/L$, $1/L^2$, and $1/L^5$ in this limit respectively (see \eqref{eq:pppot2} and \eqref{eq:appot2}).

For distances $L\gtrsim 1/m_>$, the partial potentials are exponentially suppressed for all configurations. We find that for $L\gtrsim1/m_<$, the Dirac and Majorana partial potentials have distinct $L$-dependencies with $1-V_M/V_D \sim\mathcal{O}\left(1\right)$. The latter effect occurs when the partial potential is already exponential suppressed, since $1/m_< \ge 1/m_>$. 
Hence we find that the contributions from the cross-term partial potentials ($m_i\neq m_j$) are not helpful in making a Dirac/Majorana distinction.

We find the current sensitivity to neutrino forces remains very low. 
For plates at a separation of $L\sim 1/m$ where $1-V_M/V_D\sim \mathcal{O}(1)$, we use data from a recent Casimir force experiment\footnote{This result is recast in \cite{Brax:2017xho} to bound the relevant quantum force.} \cite{Chen:2014oda} to determine that 20 orders of magnitude still remain between current experimental limits and the quantum neutrino force. 

Ref. \cite{Stadnik:2017yge} recently claimed that bounds from muonium spectroscopy place experimental limits just two orders of magnitude shy of being able to detect quantum forces from neutrinos. In their analysis, a non-relativistic formalism was used and it was assumed that the form of the potential \eqref{eq:aapot} and electronic wavefunctions are valid down to $r\sim 1/m_Z$. In an upcoming work \cite{Costantino:future}, this bound will be checked in a relativistic formalism. Unfortunately, even if the bounds in \cite{Stadnik:2017yge} hold, the confusion theorem renders a Dirac/Majorana distinction nearly impossible by this probe, as $1-V_M/V_D\sim \mathcal{O}\left(10^{-11}\right)$ for  $r_{\text{Bohr}} \sim L\ll 1/m$.

Hence for both atomic and micron scale experiments, we conclude that there are still many orders of magnitude in sensitivity needed to make a Dirac/Majorana distinction with quantum neutrino forces.

\section*{Acknowledgments}

A.C.~is supported by the National Science Foundation Graduate Research Fellowship Program under Grant No.~1840991.
  S.F.~is supported by the S\~ao Paulo Research Foundation (FAPESP) under grants \#2011/11973, \#2014/21477-2 and \#2018/11721-4, and funded in part by the Gordon and Betty Moore Foundation through a Fundamental Physics Innovation Visitor Award (Grant GBMF6210).

\appendix

\section{Lagrangians}\label{app:fermions}

Here we give more details on Lagrangians in the 2 and 4-component formalisms. 
The 2-component neutrino charged under $SU(2)_L$ is denoted $\nu_L$, the singlet neutrino is denoted $\nu_R$. 
The $L$ and $R$ labels only refer to the gauge charge.  $\nu_L$ and $\nu_R$ are left-handed  \textit{i.e.} transform as the $(1/2,0)$ representation of the Lorentz group.

The free Lagrangian for $\nu_L$ in case of Dirac and Majorana masses are given by 
\be
\mathcal{L}_{D,{\rm kin}}=  i \nu^{i \dagger}_L  \bar \sigma^\mu \partial_\mu \nu^i_L +
i \nu^{i \dagger}_R  \bar \sigma^\mu \partial_\mu \nu^i_R
- m_i \left( \nu^i_L\nu^i_R+\nu^{i\dagger}_L\nu^{i\dagger}_R\right)
\ee
\be
\mathcal{L}_{M,{\rm kin}}=i \nu^{i \dagger}_L  \bar \sigma^\mu \partial_\mu \nu^i_L - \frac{m_i}{2} \left( \nu^i_L\nu^i_L+\nu^{i\dagger}_L\nu^{i\dagger}_L\right)\,.
\ee
Integrating out the $Z$ boson in the electroweak Lagrangian gives the effective interaction
\be
{\cal L}^Z_{\rm int}=\frac{4 G_{\rm F}}{\sqrt{2}}  J_Z^\mu J_{Z\mu} \supset  \frac{4 G_{\rm F}}{\sqrt{2}}  (\nu^{i \dagger}_L \bar \sigma^\mu \nu^i_L) J_{\psi\mu}
\ee
where $J_{\psi\mu}$ is the weak neutral current for fields other than neutrinos. 
Integrating out the $W$ bosons gives
\be
{\cal L}^W_{\rm int}=\frac{8 G_{\rm F}}{\sqrt{2}}  J_W^{\mu-} J^+_{W\mu} \supset \frac{4 G_{\rm F}}{\sqrt{2}} (e^{ \dagger}_L \bar \sigma^\mu \nu^i_L) (\nu^{i \dagger}_L \bar \sigma^\mu e_L)=-
\frac{4 G_{\rm F}}{\sqrt{2}} (\nu^{i \dagger}_L \bar \sigma^\mu \nu^i_L) (e^{ \dagger}_L \bar \sigma^\mu e_L) .
\ee
We used a Fierz rearrangement in the last step.

The $\nu_L$ field can be described as a 4-component Majorana fermion
\be
\nu_M=\begin{pmatrix}\nu_L \\ \nu_L^\dagger \end{pmatrix}\label{eq:nuM}
\ee
The $\nu_L$, $\nu_R$ can be combined into a Dirac fermion 
\be
\nu_D=\begin{pmatrix}\nu_L \\ \nu_R^\dagger \end{pmatrix} \,.
\label{eq:nuR}
\ee
This provides the Dirac and Majorana fields used in our calculations. 
The neutrino bilinear in the various representations is expressed as
\be
\nu^{ \dagger}_L \bar \sigma^\mu \nu_L = 
-\frac{1}{2}\bar\nu_M \gamma^\mu \gamma_5\nu_M
=\bar\nu_D \gamma^\mu \frac{1-\gamma_5}{2}\nu_D \,.
\ee
Using this and the definitions \eqref{eq:nuM}, \eqref{eq:nuR} in ${{\cal L}_{D/M, {\rm kin} }}+{\cal L}_{\rm int}$ gives the 4-component Lagrangians \eqref{eq:lagrangian_D}, \eqref{eq:lagrangian_M}. 

In these 4-component Lagrangians, the relevant couplings to SM fermions in case of unpolarized matter are the vector ones. We find
\begin{align}
g^V_{ij}&=(1-4 s^2_w)\delta_{ij}     &{\rm if}\quad\quad  \psi&=p \\
g^V_{ij}&=-\delta_{ij}   &{\rm if}\quad\quad  \psi&=n \\
g^V_{ij}&= 2 U_{ie}U^\dagger_{e j}   -(1-4 s^2_w)\delta_{ij}    &{\rm if} \quad\quad \psi&=e .
\end{align}

\section{Point-Point Derivation}\label{se:aamath}

For this calculation, we follow the steps outlined in \cite{Brax:2017xho,Costantino:2019ixl}. 
The scattering amplitude corresponding to the loop diagram in Fig. \ref{fig:fd} is given by
\be
i\mathcal{M}_{ij} = -S^{\mu\nu}\int \frac{d^4k}{(2\pi)^4} \text{Tr}\left[\left(\frac{(\slashed{k}+m_i)\gamma_\mu(\eta-\gamma_5)}{k^2 - m_i^2 +i\varepsilon}\right) \left(\frac{(\slashed{q} + \slashed{k}+m_j)\gamma_\nu(\eta-\gamma_5)}{(q + k)^2 - m_j^2 +i\varepsilon}\right)\right]
\ee
with
\be
S^{\mu\nu}\equiv 2^{-2-\eta}G_F^2\bar{u}_{p_1'}\gamma^\mu\left(g^V_{ij}-g^A_{ij}\gamma_5\right) u_{p_1}\bar{u}_{p_2'}\gamma^\nu\left(g^V_{ji}-g^A_{ji}\gamma_5\right) u_{p_2}.
\ee
When both point sources are nonrelativistic and polarized, the spin structure simplifies to
\begin{align}
\bar{u}_{p_1'}\gamma^\mu u_{p_1}  
\approx
2m_\psi\delta_0^\mu\mathbbm{1}& \quad\quad \quad \bar{u}_{p_1'}\gamma^\mu\gamma^5 u_{p_1} 
\approx
2m_\psi\delta_a^\mu\sigma^a\label{eq:exp} \ .
\end{align}

We introduce Feynman parameters to simplify the loop integral. Upon dimensional regularization, the resulting integrals are given by
\begin{align}
\int \frac{d^4l}{(2\pi)^4}\frac{1}{(l^2 - \Delta_{ij})^2} 
&\longrightarrow
\frac{-i}{(4\pi)^2}\ln \left(\frac{\Delta_{ij}}{\Lambda^2}\right)
\\
\int \frac{d^4l}{(2\pi)^4}\frac{l^2}{(l^2 - \Delta_{ij})^2} 
&\longrightarrow
\frac{-2i\Delta_{ij}}{(4\pi)^2}\ln \left(\frac{\Delta_{ij}}{\Lambda^2}\right)
\end{align}
with $\Delta_{ij} = xm_j^2+(1-x)m_i^2 -x(1-x)q^2$. The remaining function can be decomposed into the basis of 
\begin{align}
f_{mn} &\equiv \int_{0}^{1}dx\, x^m(1-x)^n\ln \left(\frac{\Delta_{ij}}{\Lambda^2}\right) \ .\label{eq:fmn}
\end{align}
These functions have a branch cut when $\Delta_{ij}<0$. The discontinuity across this branch cut is
\be
\mathrm{D}[f_{mn}] = 2\pi i\int_{x_-}^{x_+}dx x^m(1-x)^n
\ee
for 
\be
x_\pm \equiv \frac{q^2+\left(m_i^2-m_j^2\right)\pm\sqrt{\left(q^2-(m_i-m_j)^2\right)\left(q^2-(m_i+m_j)^2\right)}}{2q^2}.
\ee

The amplitude is related to the spatial potential by
\be
V_{ij}({\bm r}) 
= \int\frac{d^3q}{(2\pi)^3} 
\frac{-\mathcal{M}_{ij}(\vec{q},q_0\approx0)}{4m_\psi^2}
e^{i\vec{q}\cdot\vec{r}}.\label{eq:potentialamplitude}
                                                  \ee
Inside the Fourier transform, we identify the transfer momentum with a gradient, $\vec{q} = -i\nabla$. This gives an expression for the potential that is a Fourier transform of a function that only depends on the magnitude $|\vec{q}|$ and the gradient.
\footnote{For more details, please see \cite{Brax:2017xho,Costantino:2019ixl}.} 
The magnitude is analytically continued as $|\vec{q}|=i\lambda$, and after some manipulations we find
\begin{align}
V_{ij}({\bm r}) 
&=
\frac{1
}{4m_\psi^2}
\int_{m_i+m_j}^{\infty}
\frac{d\lambda}{(2\pi)^2}\lambda
\,
\textrm{D}\left[i\mathcal{M}_{ij}\left( 
\lambda, -i\vec{\nabla},q_0\approx0\right)\right]
\frac{e^{-\lambda r}}{r}
\,.
\end{align}
Summing the partial potentials from three generations of neutrinos then yields \eqref{eq:aaintegral}.

\section{Casimir Force from the Path Integral \label{app:path}}

We show how to derive the potential between generic extended sources, shown in \eqref{eq:Vgen}. 
Start from an effective Lagrangian with a bilinear coupling between a Dirac fermion $\Psi$ and a nonrelativistic density of matter $J$,
\be
{\cal L}=i \bar \Psi \slashed{\partial} \Psi - m \bar \Psi \Psi + \bar \Psi \Gamma \Psi J({\bm x})\,
\ee
where $\Gamma$ can be any Lorentz structure. 

We are interested in calculating the energy of a configuration involving two  objects $J_1$, $J_2$ acting as sources, both
described by the distribution $J=J_1+J_2$. 
The relevant information is contained in the generating
functional of connected correlators $W[J]$, given by
\be
 Z[J]=\int {\cal D}\bar\Psi {\cal D}\Psi e^{i \int d^4 x  {\cal L}[\Psi,J]}= e^{-i W[J]} \,.
\ee
When the source is static, $W[J] =E[J] T$  
where $T=\int dt$ is the integral over time. $E[J]$ is the quantum vacuum energy. 
At one-loop level, the vacuum energy $E[J]$ is given by the functional determinant (see \textit{e.g.} \cite{Peskin:257493}) \begin{align}
E[J]&=i \ln {\rm Det} \left[i \slashed{\partial} - m + \Gamma J\right] \\
&= i \left(  \sum_{n=1}^{\infty} \frac{(-1)^{n+1}}{n} \left(\frac{ \Gamma J}{i \slashed{\partial} - m} \right)^n
+ {\rm Tr} \ln \left[i \slashed{\partial} - m \right] \right),
\end{align}
where ${\rm Det}$/${\rm Tr}$ is the determinant/trace in the functional sense. 

$E[J]$ contains infinities---the observable quantity is rather the variation $\partial_L E[J]$, which gives the Casimir force.  
In the limit where the $\Gamma J$ contribution can be treated perturbatively, the leading contribution to $\partial_L E[J]$ is from the $n=2$ term, 
\be
\partial_L E[J] \supset i \int d^3 {\bm x} \int d^4 x' \tr \left[ \Gamma \partial_L J({\bm x}) \Delta(x-x') \Gamma J({\bm x}') \Delta(x'-x) \right]
\ee
where $\tr$ is the trace on spinor indexes. 
The piece of potential associated to this term is found to be
\be
V(L)=i \int d^3 {\bm x} \int d^4 x' \tr \left[ \Gamma J_1({\bm x}) \Delta(x-x') \Gamma J_2({\bm x}') \Delta(x'-x) \right] \,.
\ee
Restoring the coupling constant yields \eqref{eq:Vgen} in the Dirac case.

\bibliographystyle{JHEP}
\bibliography{biblio}

\end{document}